\documentclass{amsart}
\usepackage{amsmath}
  \usepackage{graphics} %% add this and next lines if pictures should be in esp format
  \usepackage{epsfig} %For pictures: screened artwork should be set up with an 85 or 100 line screen

\newtheorem{theorem}{Theorem}[section]
\newtheorem{corollary}{Corollary}[section]

\theoremstyle{definition}

\newtheorem{remark}{Remark}[section]

\begin{document}
%% Place the running title of the paper with 40 letters or less in []
 %% and the full title of the paper in { }.
\title[Exact solutions  to ideal hydrodynamics of inelastic gases]{Exact solutions
with singularities to ideal hydrodynamics of inelastic gases}

% Place all authors' names in [ ] shown as running head;
% No more than 40 letters. Leave { } empty
% Please use `and' to connect the last two names if applicable
\author[Olga Rozanova]{}

% It is required to enter MSC and Keywords.
% Please provide minimum  5 keywords.

% Email address of each of all authors is required.
% You may list email addresses of all other authors, separately.
 \email{rozanova@mech.math.msu.su}

% Put your short thanks below. For long thanks/acknowlegements,
%please go to the last acknowlegments section.
\thanks{Supported by RFBR Project Nr. 12-01-00308
and by the government grant of the Russian Federation for support of
research projects implemented by leading scientists, Lomonosov
Moscow State University under the agreement No. 11.G34.31.0054.}

\maketitle

% Enter the first author's name and address:
\centerline{\scshape Olga Rozanova}
\medskip
{\footnotesize
% please put the address of the first author
 \centerline{Mechanics and Mathematics Faculty,
Moscow State University, Moscow, 119992, Russia}
} % Do not forget to end the {\footnotesize by the sign }

\begin{abstract}We construct a large family of exact solutions to the hyperbolic
system of 3 equations of ideal granular hydrodynamics in several
dimensions for arbitrary adiabatic index $\gamma$.  In dependence of
initial conditions these solutions can keep smoothness for all times
or develop singularity.  In particular, in the 2D case the
singularity can be formed either in a point or along a line. For
$\gamma=-1$ the problem is reduced to the system of two equations,
related to a special case of the Chaplygin gas.  In the 1D case this
system can be written in the Riemann invariant and can be treated in
a standard way. The solution to the Riemann problem in this case
demonstrate an unusual and complicated behavior.
\end{abstract}

\bigskip

% The name of the associate editor will be entered by an editorial staff
% "Communicated by the associate editor name" is not needed for special issue.
% \centerline{(Communicated by the associate editor name)}

%The abstract of your paper

%The title of your section 1
\section{Introduction}

The motion of the dilute gas  where the characteristic hydrodynamic
length scale of the flow  is sufficiently large and the viscous and
heat conduction terms can be neglected  is governed by the systems
of equations of ideal granular hydrodynamics \cite{Brilliantov}.

This system is given in ${\mathbb R}\times{\mathbb R}^n,\, n\ge 1,$
and has the following form:
\begin{equation}\label{1.1}
\partial_t \rho+{\rm div}_x (\rho v)=0,
\end{equation}
\begin{equation}\label{1.2}
\partial_t(\rho v)+{\rm Div}_x (\rho v \otimes v)=-\nabla_x
p,
\end{equation}
\begin{equation}\label{1.3}
\partial_t T+(v,\nabla_x T) + (\gamma-1) T {\rm div}_x v=-\Lambda
\rho T^{3/2},
\end{equation}
where $\rho$ is the gas density, $ v=(v_1,...,v_n)$ is the velocity,
$T$ is the temperature, $p=R\rho T$ is the pressure (the constant
$R$ is a adiabatic invariant, for the sake of simplicity we set
$R=1$), and $\gamma$ is the adiabatic index, $\Lambda=const>0.$ We
denote ${\rm Div}_x$ and ${\rm div}_x$ the divergence of tensor and
vector with respect to the space variables. The only difference
between equations (\ref{1.1})--(\ref{1.3}) and the standard ideal
gas dynamic equations (where the elastic colliding of particles is
supposed) is the presence  of the inelastic energy loss term
$-\Lambda \rho T^{3/2}$ in (\ref{1.3}).

The granular gases are now popular subject of experimental,
numerical and theoretical investigation (e.g. \cite{Brilliantov},
\cite{Ludvig}, \cite{Meerson2} and references therein).
 The Navier-Stokes granular hydrodynamics is the natural language for
 a theoretical description of granular macroscopic flows.  A characteristic feature
of  time-dependent solutions of the continuum equations is a
formation of finite-time singularities: the density blowup signals
the formation of close-packed clusters.

System (\ref{1.1}) -- (\ref{1.3}) can be written in a  hyperbolic
symmetric form in variables $\,\rho, \,v, K=p \rho^{-\gamma}$ and
the Cauchy problem
$$(\rho, v, T \rho^{1-\gamma})\big|_{t=0}=
(\rho_0, v_0, T_0 \rho_0^{1-\gamma}) $$ is locally solvable in the
class of smooth functions.
%\cite{Kato}.

System (\ref{1.1}) -- (\ref{1.3}) has no constant solution except
the trivial one $(p\equiv 0).$ Another trivial solution is
$v=p=T\equiv 0, \, \rho(t,x)=\rho_0(x)$. At the same time there
exists a solution
\begin{equation}\label{cooling} \rho, v, p = {\rm const}, \quad T =T(t)=(\frac{\Lambda \rho_0
t}{2}+T(0)^{-1/2})^{-2},
\end{equation}
 where $T(0)$ is the initial value of
temperature (the Haff's law). This solution is called the
homogeneous cooling state.

Here we are going to construct new exact solutions to the ideal
granular hydrodynamics with a concentration property and to compare
them with the known family of solution obtained earlier in
\cite{Meerson2}.

\section{Family of exact solutions in 1D\cite{Meerson2}} The authors employ Lagrangian coordinates and
derive a broad family of exact non-stationary non-self-similar
solutions. These solutions exhibit a singularity, where the density
blowups in a finite time when starting from smooth initial
conditions. Moreover, the velocity gradient also blowups while the
velocity itself and develop a cusp discontinuity (rather then a
shock) at the point of singularity.

System (\ref{1.1}) - (\ref{1.3}) in the Lagrangian coordinates takes
the form
%\begin{equation}\label{Lagrange}
 $$\frac{\partial}{\partial
t} \left(\frac{1}{\rho}\right)=\frac{\partial v}{\partial m},\qquad
\frac{\partial v}{\partial t} = - \frac{\partial p}{\partial
m},\qquad \frac{\partial p}{\partial t}=-\gamma p\rho \frac{\partial
v}{\partial m}-\Lambda p^{3/2}\rho^{1/2},
$$
%\end{equation}
and under certain assumptions can be reduced to $$\frac{\partial^2
p}{\partial m^2}=-\mu^2 p,\qquad \mu=\frac{\Lambda}{\gamma
\sqrt{2}}.$$

Here  $ m(x,t) = \int\limits_0^x \,\rho (\xi,t)\,d\xi$ is the
Lagrangian mass coordinate. The  solution is the following:
\begin{equation}
p=2A \cos (\mu m),\quad A={\rm const},\quad
\rho(m,t)=\frac{\rho(m,0)}{(1-\mu t \sqrt{A\rho(m,0)\cos \mu m})^2}.
\end{equation}
The rate of concentration at the maximum point of density as $ t\to
t_*$ is $\rho(0,t)\sim {\rm const}(t_*-t)^{-2},$ the behavior of
solution at different moments of time and formation of the
singularity is presented at Fig.1.
\begin{figure}[h]\label{m}
\begin{minipage}{0.7\columnwidth}
\centerline{\includegraphics[width=0.7\columnwidth]{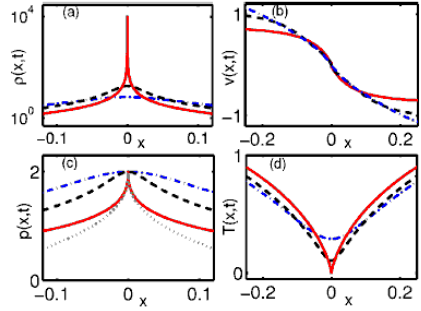}}
\end{minipage}%
\caption{Finite mass solution \cite{Meerson2}}
\end{figure}
\section{Solutions with a constraint} %utions with uniform deformation, arbitrary $\gamma$}

Let us introduce a new dependent variable $z(t,x)$ as follows:
$z=\rho-\phi(t) T^{-\frac{1}{2}}$, $\phi(t)$ is an arbitrary
differentiable function. Thus, in the variables $z, T, v$ the system
(\ref{1.1})--(\ref{1.3}) takes the form
\begin{equation}\label{1.4n}
\partial_t z+{\rm div}_x (z v)-\frac{\Lambda}{2}\, \phi(t) \, z +
(\gamma+1)\, \phi(t)\,  T^{-\frac{1}{2}}\, {\rm div}_x \, v +
(2\phi'(t)+{\Lambda}\phi^2(t))\,T^{-\frac{1}{2}} =0,
\end{equation}
\begin{equation}\label{1.5n}
\partial_t v+ (v, \nabla_x)\, v =-\frac{1}{z+\phi(t) T^{-\frac{1}{2}}}\,\nabla_x
(zT+\phi(t)T^{\frac{1}{2}}),
\end{equation}
\begin{equation}\label{1.6n}
\partial_t T+(v,\nabla_x T) + (\gamma-1) \,T\, {\rm div}_x\, v=-\Lambda
z\, T^{3/2}- \Lambda \,\phi(t)\, T.
\end{equation}

We consider a particular class of solutions characterized by
property $z=0.$ There are two possibilities:
\begin{itemize}
\item $\gamma=-1$ and $\phi(t)$ is a solution to ODE that can be immediately solved:
\begin{equation}\label{1.7n}
\phi'(t)=-\frac{\Lambda}{2}\phi^2(t), \quad
\phi(t)=\left(\frac{\Lambda}{2} t + \rho^{-1}_0(x)
T_0^{-\frac{1}{2}}(x)\right)^{-1},
\end{equation}

\item $v(t,x)=\alpha(t)x+\beta(t)$ and $\phi(t)$ is a solution to
ODE
\begin{equation}\label{1.8n}
\phi'(t)= -\frac{\gamma+1}{2}\, \phi(t)\, {\rm tr}\, \alpha(t) -
\frac{\Lambda}{2}\phi^2(t).
\end{equation}
\end{itemize}

The first possibility is the case of Chaplygin-like \cite{Chaplygin}
gas,
%\cite{Chaplygin},
where the state equation is chosen as
\begin{equation}\label{1.9n}
p=p_0- \rho^{-1}, \quad p_0=const >0.
\end{equation}
 The system
(\ref{1.1})--(\ref{1.3}),(\ref{1.9n}) with the constraint $z=0$ can
be  reduced to a couple of equations
\begin{equation}\label{1.10n}
\partial_t v+ (v, \nabla_x)\, v ={ T^{\frac{1}{2}}}\,\nabla_x
(T^{\frac{1}{2}}),
\end{equation}
\begin{equation}\label{1.11n}
\partial_t T^{\frac12}+(v,\,\nabla_x T^{\frac12}) - T^{\frac12} {\rm div}_x\,v=- \frac{\Lambda}{2} \,\phi(t)\,
T^{\frac12},
\end{equation}
where $\phi(t)$ is given by (\ref{1.7n}).

In the 1D case this system as any system of two equations can be
written in the Riemann invariants, this allows to apply the
technique usual for  gas dynamics, we will do this in
Sec.\ref{Chaplygin}.

In the second case, for an arbitrary $\gamma$ the equation
(\ref{1.4n}) can be satisfied only for $v(t,x)=\alpha(t)x+\beta(t)$,
where $\phi(t)$ solves (\ref{1.8n}). This case will be considered in
Sec.\ref{Uniform}.

\subsection{Solutions with uniform deformation, arbitrary
$\gamma$}\label{Uniform}

It is known that for usual gas dynamics equations the solutions with
linear profile of velocity $v(t,x)=\alpha(t)x+\beta(t)$, where
$\alpha(t)$ is a matrix $n\times n$ and $\beta(t)$ is an $n$ -
vector, $x$ is a radius-vector of point, constitute a very important
class of solutions \cite{Sedov}. For the system of granular
hydrodynamics these solutions give a possibility to construct a
singularity arising from initial data.

First of all from (\ref{1.10n}, (\ref{1.11n}) we get that in this
case $T$ has to solve the system
\begin{equation}\label{1.12n}
(\partial_t \alpha(t)+\alpha^2(t))x + (\partial_t
\beta(t)+\alpha(t)\beta(t)) =-\frac12 \,\nabla_x T,
\end{equation}
\begin{equation}\label{1.13n}
\partial_t T+((\alpha x+\beta),\nabla_x T) + (\gamma-1) T \,{\rm tr}\,\alpha(t) =- \Lambda \,\phi(t)\,
T,
\end{equation}
and the structure of the field of velocity requires a special
structure of the field of temperature, namely,
\begin{equation}\label{1.14n}
T(t,x)=x^T\,A(t)\, x + (B(t),x) +C(t).
\end{equation}
Thus, we get a system of $\frac{3 n^2 + 5n +4}{2}$ nonlinear
differential equations for components of the square  matrix
$\alpha(t)$, the square symmetric matrix  $A(t)$, vectors $\beta(t)$
and $B(t)$, the scalar functions $C(t)$ and $\phi(t)$, namely
\begin{equation}\label{alpha}
\alpha'(t)+\alpha^2(t)+A(t)=0,\qquad
%\end{equation}
%\begin{equation}\label{beta}
\beta'(t)+2\alpha(t)\beta(t)+\frac{1}{2}B(t)=0,
\end{equation}
\begin{equation}\label{A}
A'(t)+2A(t)\alpha(t)+(\gamma-1){\rm
tr}\,\alpha(t)A(t)+\Lambda\phi(t) A(t)=0,
\end{equation}
\begin{equation}\label{B}
B'(t)+2A(t)\beta(t)+B(t)\alpha(t)+(\gamma-1){\rm
tr}\,\alpha(t)B(t)+\Lambda\phi(t) B(t)=0,
\end{equation}
\begin{equation}\label{C}
C'(t)+(B(t),\beta(t))+(\gamma-1){\rm
tr}\,\alpha(t)C(t)+\Lambda\phi(t) C(t)=0,
\end{equation}
and (\ref{1.8n}). This system can be explicitly (in the simplest
cases) or numerically integrated, one can study its qualitative
behavior. The component  of density can be found as
\begin{equation}\label{1.15n}  \rho(t,x)=\frac{\phi(t)}{(x^T\,A(t)\, x + (B(t),x)
+C(t))^{1/2}},
\end{equation}
$\rho(t,x)\sim const*|x-x_0|^{-1}$ in the point $x_0$ of the
singularity formation. Therefore the singularity is integrable for
$n>1$. Nevertheless, the total mass is infinite for this solution,
since $\int\limits_{{\mathbb R}^n}\rho \, d x$ diverges as $|x|\to
\infty$.

Let us consider the simplest non-rotational case: $A(t)=a(t) \mathbb
I$, $\alpha(t)=\alpha_1(t){\mathbb I}$
 $B(t)=0, $ $\beta(t)=0,$
 where $\mathbb I$ is the unit matrix.
 The system above comes  to 4 equations:
 \begin{equation}\label{phi_n}
\phi'(t)+\frac{n}{2}(\gamma+1)\, \phi(t)\, \alpha_1(t) -
\frac{\Lambda}{2}\phi^2(t),\qquad
%\end{equation}
%\begin{equation}\label{alpha1}
\alpha_1'(t)+\alpha_1^2(t)+a(t)=0,
\end{equation}
\begin{equation}\label{a}
a'(t)+((2+n(\gamma-1))\alpha_1(t)+\Lambda\phi(t)) a(t)=0,
\end{equation}
\begin{equation}\label{á}
C'(t)+(n(\gamma-1)\alpha_1(t)+\Lambda\phi(t)) C(t)=0.
\end{equation}

We are going to find  asymptotics of the solution at the point
$t=t_*>0$ of the  singularity appearance.

Systems (\ref{phi_n})-(\ref{á}) is a polynomial system
\begin{equation}\label{system_f}
\dot {\bf x} = {\bf f} ({\bf x}),\quad {\bf f} : {\mathbb R}^k \to
{\mathbb R}^k,\end{equation} and we can study the occurrence of
blow-up analyzing the solutions locally around their movable
singularities using a set of methods based on the construction of
local series. Following \cite{Goriely_Hyde_JDE}, \cite{Goriely},  we
build local series ($\Psi$-series) of the form:
$$
 {\bf x} = \Psi(\lambda, s, t)= \lambda \tau^s(1 + h(\tau, {\rm ln} \tau)),
$$
where $\tau =  t_*-t$ and $h(\tau; {\rm ln}\tau )$ is a power series
in its argument which vanishes as $\tau \to 0$. The notation
$\lambda \tau^s $ refers to the vector whose $i$-th component is
$\lambda_i \tau^{s_i} .$   In order to obtain the leading behavior
$\lambda \tau^s $ of the solution around $t_*$ we look for all {\em
negatively quasihomogeneous truncations} $ \bf \hat f$ of the vector
field ${\bf f} =\bf\hat f + \breve{f}$ such that the {\em dominant
behavior} ${\bf x} =\lambda \tau^s, \quad \lambda \in {\mathbb C}^k$
is an exact solution of the truncated system ${\bf \dot x} = \bf\hat
f (x)$ and
$${\bf \breve{f}}(\lambda \tau^s) \sim  \breve{\lambda} \tau^{s +
\breve{s}-1}, \quad \breve{s}\in {\mathbb Q}^k, \quad \breve{s}_i>0,
$$
as $\tau\to 0.$ Each truncation defines a {\em dominant balance}
$(\lambda, s)$ and every balance corresponds to the first  term
$\lambda \tau^s$ in an expansion around movable singularities. For
such an expansion to describe a general solution, the $\Psi$-series
must contain $k -1$ arbitrary constants in addition to the arbitrary
parameter $t_*.$ The position in the power series where these
arbitrary constants appear is given by the {\em resonances}.  They
are given by the eigenvalues of the matrix $R$:
$$
R =- D \hat{\bf f}(\lambda) -  diag(s),$$ where $D \hat{\bf
f}(\lambda)$ is the Jacobian matrix evaluated on $\lambda$. The
resonances are labeled $r_i, \,i = 1,...,k$ with $r_1=-1$. Each
balance defines a new set of resonances.

\begin{theorem}\label{G_book}(\cite{Goriely}, T.3.8)
Consider a real analytic system (\ref{system_f})and assume that it
has a balance $(\lambda, s)$  such that $r_j
>0 $   for all $j > k-m+1,\,1\le m\le k$ and $\lambda \in {\mathbb R}^n$. Then
there exists a $m$-dimensional manifold $S^m_0 \subseteq {\mathbb
R}^k$ of initial conditions leading to a finite
 time blow-up,
 for all $x_0 \in S^m_0$, i.e. there exists $t_* \in \mathbb R_+$ for
which $|{\bf x}(t; x_0)|\to \infty $ as $t\to t_*$.
\end{theorem}

\begin{corollary} If  $n\ge 2$ and $\gamma>-1+\frac{2}{n}$, then there exists an open
set $\Omega$ of initial data $x_0=(\phi(0), \alpha_1(0), a(0),
C(0))$  such that the components $\phi(t)$ and $\alpha_1(t)$ of
solution to the system (\ref{phi_n})--(\ref{á})
 blow up within a finite
 time
 for all $x_0\in \Omega$.
\end{corollary}

\proof

To find main terms of asymptotic at the point of singularity we
consider a negatively quasihomogeneous truncation of the system
(\ref{phi_n})--(\ref{á}), namely
\begin{equation}\label{phi_nn}
\phi'(t)= -\frac{n}{2}(\gamma+1)\, \phi(t)\, \alpha_1(t) -
\frac{\Lambda}{2}\phi^2(t),
\end{equation}
\begin{equation}\label{alpha1a}
\alpha_1'(t)=-\alpha_1^2(t),
\end{equation}
\begin{equation}\label{a}
a'(t)=-(2+n(\gamma-1))\alpha_1(t)a(t)-\Lambda\phi(t) a(t),
\end{equation}
\begin{equation}\label{C}
C'(t)=-n(\gamma-1)\alpha_1(t)C(t)-\Lambda\phi(t) C(t).
\end{equation}

The solution to the above system is the following:
\begin{equation}\label{asimpt}
\phi(t)=-\frac{n(\gamma+1)-2}{\Lambda}(t_*-t)^{-1},\quad
\alpha_1(t)= -(t_*-t)^{-1},
\end{equation}
 $$ A(t)=A_0 (t_*-t)^{2(n-2)},\quad
C(t)=C_0 (t_*-t)^{2(n-1)},\quad
 \quad A_0, C_0=\rm
const.,
$$
$$s=diag(-1,-1,2(n-2), 2(n-1)),\quad \lambda=(-1,-\frac{n(\gamma+1-2)}{\Lambda}\, A_0,
\,C_0).$$

%In 2D in the case of vortex solution we get
%$$
%\alpha_2(t)=\alpha_2^0 (t-t_*)^{-2}, \quad \alpha_2^0=\rm const,
%$$

The resonances, computed for this balance are
$(n(\gamma+1)-2,-1,0,0).$ Theorem \ref{G_book} result that there
exists a manyfold $ S^2_0\in {\mathbb R}^n$ such that for $\phi(0),
\alpha_1(0)$ the respective solution to (\ref{phi_n}) blows up and
has (\ref{asimpt}) as a main term of asymptotics. Other components
of solution to (\ref{phi_n})-- (\ref{á}) can be found from linear
with respect to $a(t)$ and $C(t)$ equations (\ref{a}) and (\ref{C})
for any initial data (it makes sense to consider $a(0)>0, C(0)>0$).
$\square$

\begin{remark}The rate of growth of the maximum of the density as $t\to
t_*$ is $\rho(t,0)\sim {\rm const} (t-t_*)^{-n}.$
\end{remark}
\begin{remark}
If $t_*<0$, then an analogous consideration shows that there exists an open set of initial data
such that the solution to  system (\ref{phi_n})--(\ref{á})remains
bounded for all $t>0$.
\end{remark}

\begin{theorem} It $n=1$, then for any $t_*>0$ there exists a family
of exact solutions to system (\ref{phi_n})--  (\ref{á}) depending on
parameters $(\alpha_0, C_0),$ blowing up as $t\to t_*.$ This family
is physically reasonable for $ \, \alpha_0\in
(-1,-\frac{2}{\gamma+1}),\quad C_0>0, \gamma>1.$ For these solutions
the  the maximum of  density has the asymptotics $\rho(t,0)\sim {\rm
const} (t-t_*)^{\alpha_0}$ as $t\to t_*$.
\end{theorem}
\proof It can be readily checked that the balance
$s=(-1,-1,-2,-2(\alpha_0+1)),$
$\lambda=(-(2+(\gamma+1)\alpha_0)/\Lambda, \alpha_0,
-\alpha_0(\alpha_0+1),C_0)$ gives an exact solution. The restriction
on the parameters $\alpha_0$ and $C_0$ follows from the positivity
of the expression under the square root in (\ref{1.15n}) and the
positivity of $\phi(0)$. $\square$

\begin{remark} The maximum of  density as $t\to
t_*$ grows  slower than for the solution obtained in
\cite{Meerson2}.
\end{remark}

 Fig.2a presents the results of numerical computations in 2D based on system (\ref{alpha}) --(\ref{C}). The initial density
 has
the form (\ref{1.15n}). Fig.2b shows the density near the blow-up
time for $\alpha_{11}=\alpha_{22}<0,$ $\alpha_{12}=\alpha_{21}=0$,
with a concentration in a point. Fig.2c shows the density near the
blow-up time  for $\alpha_{11}<0, $
$\alpha_{22}=\alpha_{12}=\alpha_{21}=0$ with a concentration along a
line.  The computations demonstrate a complicated behavior of
solution. In particular, a vorticity can prevent the singularity
formation.

\begin{figure}[h]
\begin{minipage}{0.33\columnwidth}
\centerline{\includegraphics[width=1.2\columnwidth]{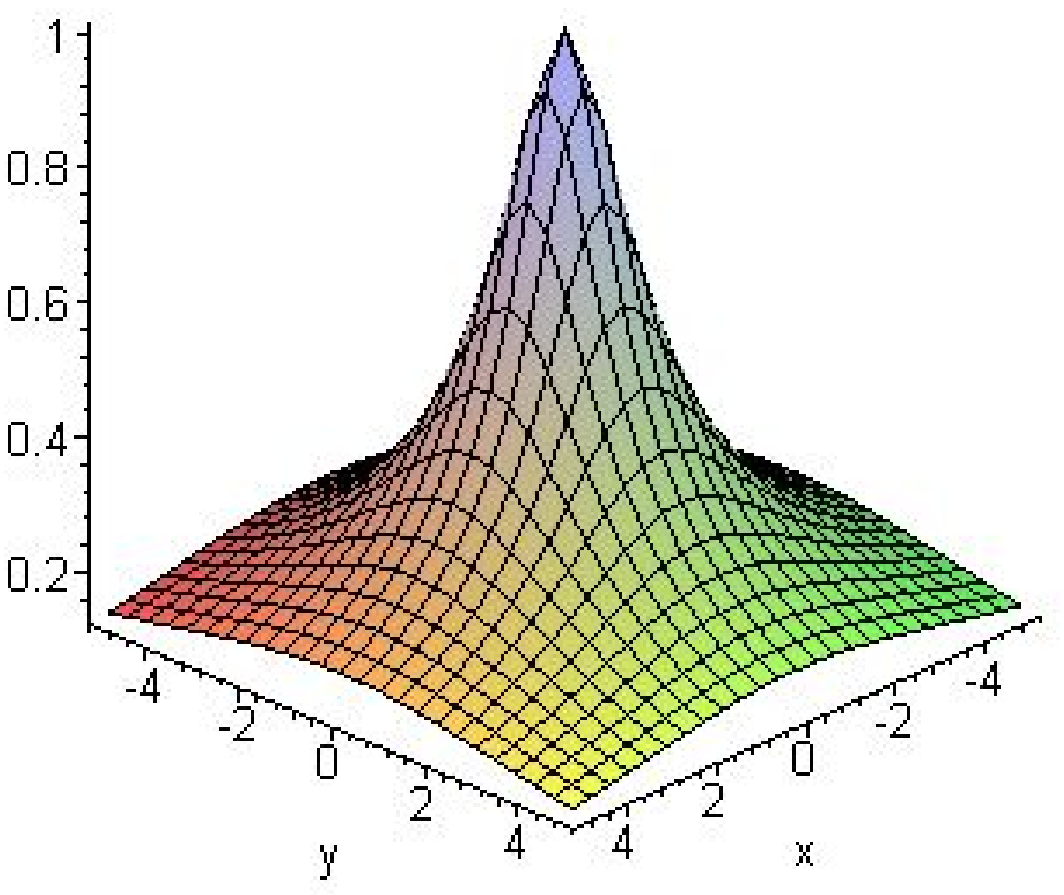}}
%\caption{}
\end{minipage}%
%\end{figure}%
%\begin{figure}[h]
\begin{minipage}{0.33\columnwidth}
\centerline{\includegraphics[width=1.2\columnwidth]{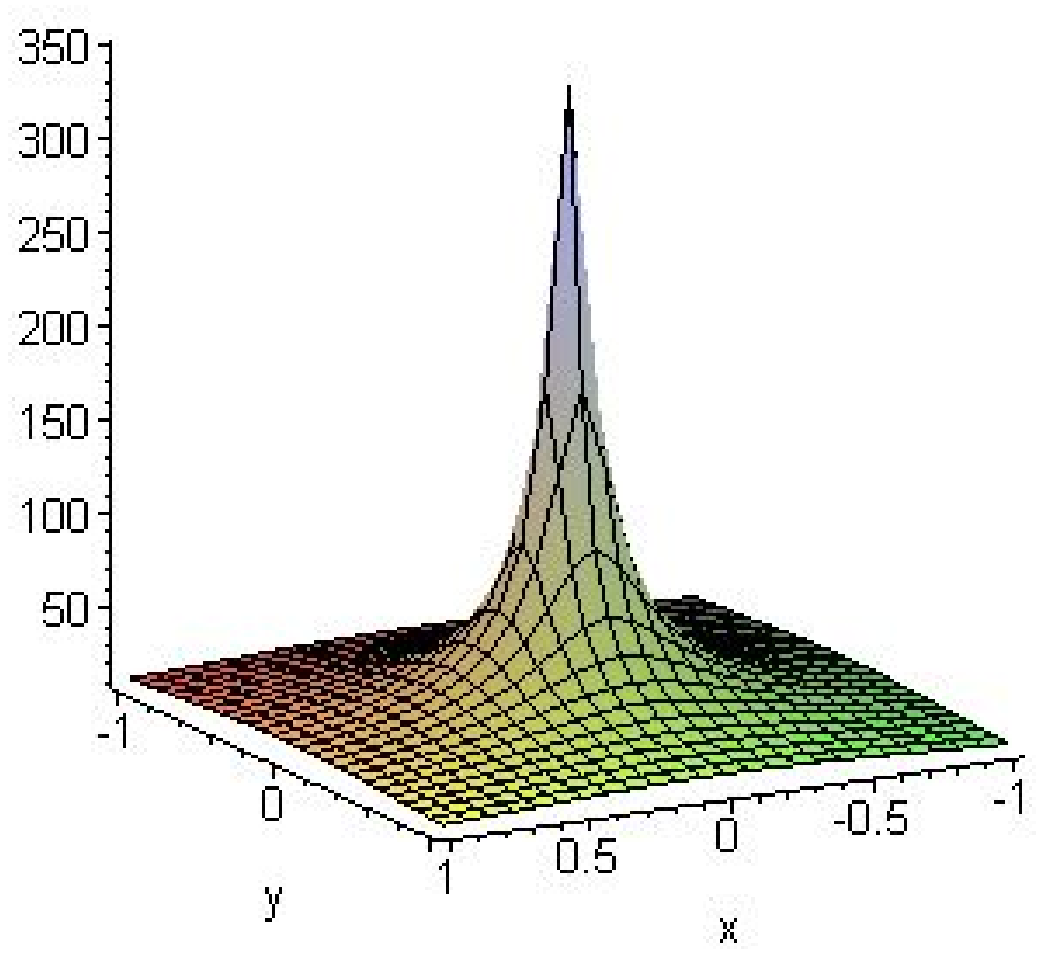}}
%\caption{}
\end{minipage}
%\end{figure}%
%\begin{figure}[h]
\begin{minipage}{0.33\columnwidth}
\centerline{\includegraphics[width=1.2\columnwidth]{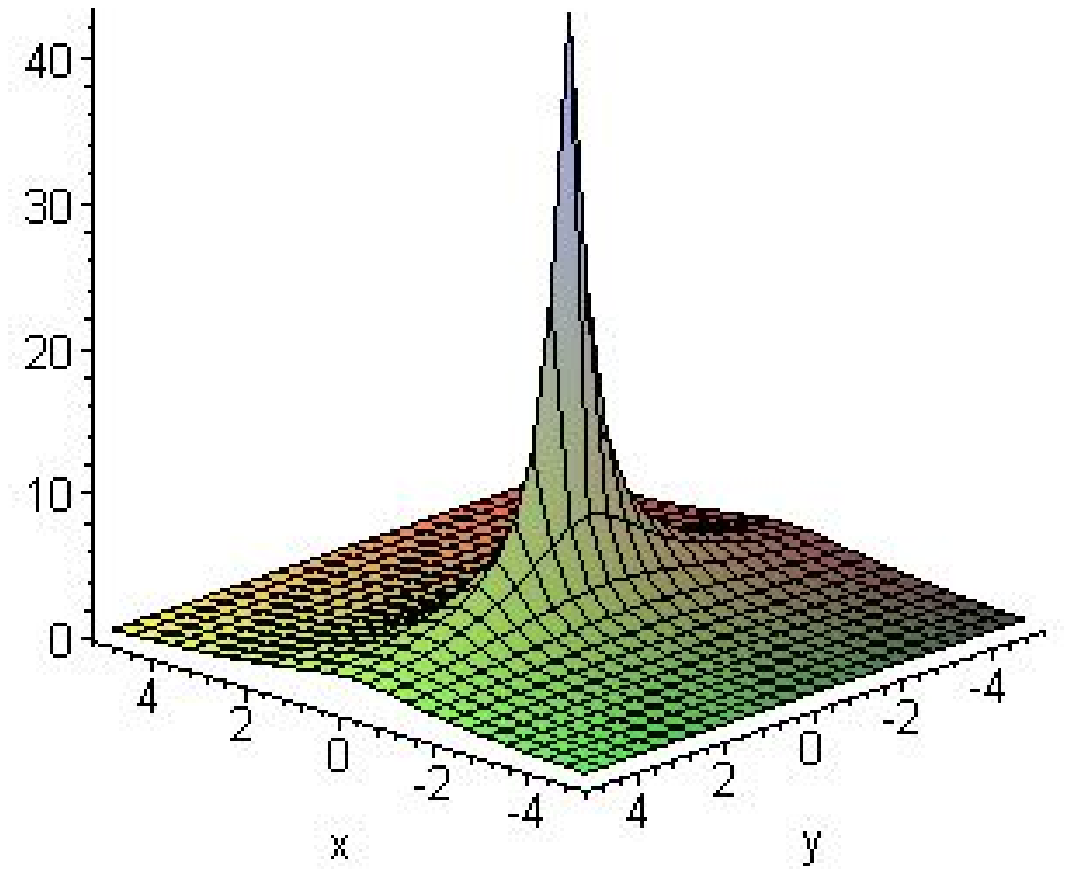}}
%\caption{,,}
\end{minipage}%
\caption{a,b,c}
\end{figure}%

\subsection{Chaplygin gas,  $n=1$.}\label{Chaplygin}

The model of gas dynamics with the pressure  given by (\ref{1.9n})
is known as the Chaplygin gas.  The Chaplygin gas is now considered
as a possible model for dark matter-energy  \cite{dark_matter_PR}.
This system can also be seen as the one-dimensional version of the
Born-Infeld system, a non linear modification of the Maxwell
equations, designed by Born and Infeld in 1934 to solve the
electrostatic divergence generated by point particles in classical
electrodynamics.% (see \cite{electridynamics} for a discussion).
 The Chaplygin system is known to be hyperbolic, linearly degenerate,
weakly stable \cite{Serre_book}.
%The sound speed is $c =
%\frac{A}{\rho}$.
 Recently this system attracted a lot of attention, e.g
\cite{Brenier}, \cite{Serre}.

The system (\ref{1.10n}), (\ref{1.11n}) can be reduced to
\begin{equation}\label{chapl_1}
\partial_t \rho+{\rm div}_x (\rho v)=0,\qquad
%\end{equation}
%\begin{equation}
%\label{chapl_2}
\partial_t (\rho v) + {\rm Div}_x (\rho v\otimes v -\frac{\phi^2(t)}{\rho})=0,
\end{equation}
recall that $\rho=\phi(t) T^{-\frac12}$. It is similar to the
Chaplygin gas system, the only difference is in the known multiplier
$\phi(t)$, for the Chaplygin gas $\phi=const$. For $n=1$
(\ref{chapl_1}) can be written in the Riemann invariants  as
\begin{equation}\label{rim_inv_s}
\partial_t s+ r \partial_x s =\frac{\Lambda \phi(t)}{4}(r-s),\qquad
%\end{equation}
%\begin{equation}\label{rim_inv_r}
\partial_t r+ s \partial_x r=\frac{\Lambda \phi(t)}{4}(s-r),
\end{equation}
where $s=v-T^{\frac12}$, $r=v+T^{\frac12}$, $\phi(t)$ is given by
(\ref{1.7n}). This system is linear degenerate, therefore provided
the solution is bounded there is no possibility for the gradient
catastrophe.

%One can construct a large family of self-similar solution $s(\xi),
%r(\xi),\,\xi=\frac{x}{\frac{\Lambda}{2}t+c}.$

\begin{theorem} The solution to the
 Riemann problem for (\ref{chapl_1}) ((\ref{rim_inv_s}))  in 1D with
 data
%(\ref{chapl_2}):
\begin{equation}\label{RP}
(v, T) = \left\{\begin{array}{rr}(v_L, T_L),& x< 0,\\(v_R, T_R),&
x>0,  \end{array}\right.
\end{equation}
in the case
\begin{equation}\label{condition_v}v_L>v_R
\end{equation}
contains  a $\delta$-singularity in the component of density. If
\begin{equation}\label{condition}
v_L\ge v_R + T_L^{\frac12}+T_R^{\frac12},
\end{equation}
then the $\delta$-singularity formation begins from the initial
moment of time.
\end{theorem}

\proof

Since the system is linear degenerate, the jumps are contact
discontinuities and move along characteristics. The solution is
based on the cooling state (\ref{cooling}). If
\begin{equation}\label{condition}
v_L< v_R + T_L^{\frac12}+T_R^{\frac12},
\end{equation}
 then the solution is
 \begin{equation}\label{sol_RP}
(v, T) = \left\{\begin{array}{cc}(v_L,T_L(t)),&
x< x_-(t),\\
(v_M(t), T_M(t)),&x_-(t)<x< x_+(t),
\\(v_R,T_L(t) ),& x> x_+(t),
\end{array}\right.
\end{equation}
with $c=\phi^{-1}(0),$  $T_L(t)=\frac{c^2 T_L}
{(\frac{\Lambda}{2}t+c)^2},$ $T_R(t)=\frac{c^2 T_R}
{(\frac{\Lambda}{2}t+c)^2},$
$$x_-(t)=v_L t-\frac{2 c T_L^{\frac12}}{\Lambda} \ln
(\frac{\Lambda}{2c}t+1)\quad x_+(t)=v_R t+\frac{2 c
T_R^{\frac12}}{\Lambda} \ln (\frac{\Lambda}{2c}t+1),$$
$$
v_M(t)=\frac{v_L+v_R+c(T_R^{\frac12} -
T_L^{\frac12})(\frac{\Lambda}{2}t+c)^{-1}}{2},$$$$
T_M^{\frac12}(t)=\frac{v_R-v_L+c(T_R^{\frac12} +
T_L^{\frac12})(\frac{\Lambda}{2}t+c)^{-1}}{2}.
$$
If $v_L\le v_R$, then $x_-(t)<x_+(t)$ for all $t>0$ and the solution
to the Riemann problem is given by (\ref{sol_RP}). If $v_L< v_R$,
then there exists a moment $t_*>0$ such that $x_-(t_*)=x_+(t_*)$.
Moreover, in the moment $t_{**}>0$ the component $T_M$ vanishes and
$t_{**}<t_*$. Thus, we have to construct a new solution starting
from $t_{**}$.

We are going to introduce a $\delta$-singularity in the density
concentrated  on the jump analogously to \cite{Brenier}. To find a
$\delta$-type singularity solution we have to use the system in its
conservative form (\ref{chapl_1}). Let us denote $x_*(t)$ the
position of the singularity and look for a solution in the form:
\begin{equation}\label{delta}
\rho(t,x)=\rho_-+[\rho] H(x-x_*(t))+ \theta (t)\delta
(x-x_*(t)),\end{equation}
\begin{equation}
 \rho(t,x) v
(t,x)=\rho v_-+[\rho v] H(x-x_*(t))+ \psi(t)\delta (x-x_*(t)),
\end{equation}
\begin{equation}
 \rho(t,x) v^2 (t,x)=\rho v_-^2-+[\rho v^2]
H(x-x_*(t))+ \Psi(t)\delta
(x-x_*(t)),\end{equation}\begin{equation}\label{delta1}  \tau(t,x)
=\tau_-+[\tau] H(x-x_*(t)),\quad \tau = \rho^{-1},
\end{equation}
$[f]=f_+-f_-,$ $f_+$ and $f_-$ are the limits of an arbitrary
function from the right and from the left side of $x_*(t)$,
respectively, $H$ is the Heaviside function. From (\ref{chapl_1}) we
get
\begin{equation}
\label{delta1x} x_*(t)=\frac{[\rho
v]t_1+\theta(t)}{[\rho]}+x_0,\end{equation}
\begin{equation}
\theta(t)=\sqrt{\left([\rho v]^2-[\rho][\rho
v^2]\right)t_1^2+\frac{4}{\Lambda}[\rho][\tau]\left(\phi(t_0)t_1-\frac{2}{\Lambda}\ln\left(\frac{\Lambda
\phi(t_0)}{2}t_1+1\right) \right)},
\end{equation}
$t_1=t-t_0$, $t_0$ and $x_0$ are the moment and the coordinate of
the $\delta$-singularity formation. Expanding the expression under
the square root at $t=t_0$ we find  the necessary condition for the
beginning of the concentration processes:
\begin{equation}\label{concentration}
[v]^2\ge [T^{1/2}(t_0)]^2.
\end{equation}
If the initial data are such that the inequality opposite to
(\ref{condition}) holds, i.e. $v_L\ge v_R+
T_L^{1/2}(0)+T_R^{1/2}(0)$, then $v_L>v_R$ and
$$
(v_L-v_R)^2\ge(T_L^{1/2}(0)+T_R^{1/2}(0))^2\ge
(T_R^{1/2}(0)-T_L^{1/2}(0))^2,
$$
such that the condition (\ref{concentration}) is satisfied. Since in
this case the solution consisting of two contact discontinuities is
impossible, the only reasonable solution is given by (\ref{delta})--
(\ref{delta1}), (\ref{delta1x}).

If $u_L>u_R$, we can define again the solution of form
(\ref{delta})-- (\ref{delta1}), (\ref{delta1x}). In principle,
initially the condition (\ref{concentration}) may fail and this
solution can exist beginning from some $\hat t, \,0< \hat t \le
t_{**}$, nevertheless, this moment $\hat t$  always exists. Further,
at least for $t>t_{*}$ this solution this solution is stable. Let us
extend this solution back  $t_{**}<t<t_{*}$. This means the
assumption that the segment $[x_-(t), x_+(t)]$ shrinks into the
point $x_*(t)$ at the moment $t_{**}$. It can be shown that the
velocity of the singular front $\dot x_*(t)\to \frac{v_L+v_R}{2}$ as
$t\to\infty$. $\square$

\begin{remark}
The question on uniqueness of the solution after the "shrinking" is
open.
\end{remark}
\begin{remark}
In \cite{Roz_Non} for any spatial dimensions a simple family of
solutions to the system (\ref{1.1}) -- (\ref{1.3}) having a
singularity in the density whereas other components are continuous
is constructed. Moreover, a family of self-similar solutions in 1D
was found.
\end{remark}

\end{document}